\documentclass[11pt,a4wide]{article}
\usepackage{newlfont}
\usepackage{amsmath}
\usepackage{graphics}
\usepackage{float}
\usepackage{graphicx}
\usepackage{epsfig}
\usepackage{multirow}
\usepackage{multicol}
\usepackage{verbatim}
\usepackage{pdflscape}
\usepackage{array}
\usepackage{hyperref}

\begin{document}

\title{Temperature dependence on Spectrum of Heavy Hybrid Mesons}
\author{Ali Zeeshan$^\ddag$, Nosheen Akbar$^\dag$ \thanks{e mail:nosheenakbar@cuilahore.edu.pk}, Sadia Arshad$^\ddag$, Ali Akgül$^\S$  \\
\textit{$\dag$ Department of Physics, COMSATS University Islamabad, Lahore Campus,}\\
{Lahore, Pakistan.}\\
\textit{$\ddag$ Department of Mathematics, COMSATS University Islamabad, Lahore Campus,}\\
{Lahore, Pakistan.}\\
\textit{$\S$  Department of Mathematics, Arts and Science Faculty, Siirt University, Siirt 56100, Turkey}}
\date{}
\maketitle

\baselineskip=0.30in
\section*{Abstract}
In this work, temperature dependence on the masses of conventional and hybrid heavy quarkonium systems is investigated. For this, a thermally screened interaction is incorporated through Debye mass $(m_{D}(T))$ into the potential models of conventional and hybrid charmonium ($c\overline{c}$) and bottomonium ($b\overline{b}$) mesons (conventional and hybrid). Mass eigenvalues for S, P, D states of these mesons are computed by power series expansion method at different values of debye mass. Comparison with available lattice-QCD-inspired potentials and previous numerical studies show strong agreement and validate the efficiency of power-series technique for calculations of mass of heavy quarkonium at finite-temperature. Our results can be helpful to explore the recent experimentally determined states of charmonium and bottomonium.

\section*{1. Introduction}
A lot of charmonium-like and bottomonium-like states are observed by experimentalists at Belle, LHCb, CDF, and BESIII. The study of these systems is most important in understanding the Quantum chromodynamics (QCD). A comprehensive study of conventional and hybrid quarkonium systems can not be possible without the detailed information about the mass spectrum and radial wavefunctions, as these quantities are further used in finding the other observed features like decay constants, radiative and electromagnetic decay widths, branching ratios etc. To get detailed information about the internal structure of these experimentally observed states, a variety of theoretical approaches including lattice QCD \cite{Liu12,Lacock97,Chen21,Dudek13}, the flux-tube model \cite{Burns}, QCD string models \cite{Inyang,khokha16} constituent-gluon models \cite{Chen22}, and QCD sum-rule techniques \cite{Chen22,Balitsky,Chen13,Chen,Chen14} have been employed.

A variety of potential models such as Yukawa potential \cite{EInyang2022}, Trigonometric Rosen-Morse potential (TRMP) \cite{MShady2022}, Cornell potential (linear plus coloumbic) \cite{Nosheen11,Nosheen17, Nosheen19, Nosheen2014,Eichen,Kuchin,Ghalenovi}, Cornell plus inverse quadratic potential \cite{Ahmadov}, Killingbeck potential (Cornell potential where the quadratic term is combined) \cite{vrscay,Killingbeck, Eichen78}, and Killingbeck potential with addition of inverse quadratic term \cite{Killingbeck80}, Cornell plus and plus inverse quadratic potential \cite{MShady2022}, Hulthén plus Hellmann potential \cite{Innying2021} are used to describe the quarkonium systems. In Ref.\cite{Innying2021}, potential is made temperature-dependent by replacing the screening parameter with Debye mass($m_D(T)$).
 This temperature dependent potential model is used to calculate the masses of heavy quarkonium by a semi-analytic approach named as power series expansion method. In recent paper, work of Ref.\cite{Innying2021} is extended for hybrid mesons to calculate their mass spectrum. For this purpose, temperature dependent Hulthén plus Hellmann potential is modified by adding an additional term, ($A (1- C r^2))$, that appear because of the excitation of the gluonic field in hybrids. Earlier, power series method is observed in Ref. \cite{2024Akber} to compute the spectrum of hybrid mesons by ignoring the temperature dependence.

The paper is arranged in such a way: In section 2, the radial Schrodinger equation (SE) is defined along with the description of potential models for conventional and hybrid mesons. The solution of SE for conventional(CM) and hybrid mesons(HM) are defined in the form of power series in section 3. The energy expressions for CM and HM are also developed in present section. Section 4 tells us about the method for calculating the parameters that are observed in potential models for determining the masses of charmonium and bottomonium systems. Results are discussed in Section 5; while concluding remarks are presented in section 6

\section* {2. Radial Schrodinger Equation}
Radial SE can be described as:
\begin{equation}
\bigg[\frac{d^{2}}{dr^{2}}+\frac{2}{r}\frac{d}{dr}-\frac{<K^2_{q \overline{q}}>}{r^{2}}+2\mu(E-U(r))\bigg]Q(r)=0, \label{E1}
\end{equation}
Here $r$ denotes the distance between quark-antiquark, $\mu$ is the reduced mass of the interacting particles (quark-antiquark), and can be expressed as $\mu=\frac{m_{q}m_{\overline{q}}}{m_{q}+m_{\overline{q}}}$ where $m_{q}$ and $m_{\overline{q}}$ are the quark and anti-quark masses respectively.
$\left\langle K_{q\overline{q}}^{2}\right\rangle =K(K+1)-2\Lambda^{2}+\left\langle J_{g}^{2}\right\rangle$, where $K$ is the angular momentum quantum number for the pair of quark anti-quark . $J_{g}$ is the angular momentum due to gluonic field and $\Lambda$ is the projection of $J_{g}$. In case of conventional mesons with gluonic field in the ground state, $\Lambda$ and $\left\langle J_{g}^{2}\right\rangle$ are zero. For hybrid mesons with gluonic field in the first excited state, the value of $\Lambda$ is one and the value of $\left\langle J_{g}^{2}\right\rangle$ is 2 \cite{Juge1999}. This show that $\left\langle K_{q\overline{q}}^{2}\right\rangle =K(K+1)$ for the mesons (with ground state gluonic field) and hybrid mesons (with gluonic field in the first excited state). The quantity $E$ denotes the energy eigenvalues associated with the radial wave functions $Q(r)$, which are discussed in detail in Section 3. The term $U(r)$ represents the interaction potential governing the quark–antiquark system.

\subsection*{2.1. Temperature dependent potential model for conventional mesons}
The potential model for conventional mesons is taken in the following form \cite{Innying2021}:
\begin{equation}
U(r)= \frac{A_{0}e^{-ar}}{1-e^{-ar}}-\frac{A_{1}}{r}+\frac{A_{2}e^{-ar}}{r}
\end{equation}
where $A_{0}$, $A_{1}$ and $A_{2}$ are potential strength constants and $a$ is the screening parameter. To make eq.(2) temperature dependent, parameter $a$ is replaced with Debye mass $(m_{D}(T))$, so the temperature dependent potential can be written as:
\begin{equation}
U(r, T)= \frac{A_{0}e^{-(m_{D}(T))r}}{1-e^{-(m_{D}(T))r}}-\frac{A_{1}}{r}+\frac{A_{2}e^{-(m_{D}(T))r}}{r}
\end{equation}
After the expansion of exponential term, the potential model in the above equation is written as:
\begin{equation}
U(r,T)= \frac{-\beta_{0}}{r}+\beta_{1}r-\beta_{2} r^{2} + \beta_{3} \label{pot}.
\end{equation}
with
\begin{multline}
-\beta_{0}= -A_{1}+A_{2}-\frac{A_{1}}{(m_{D}(T))},  \beta_{1}= \frac{A_{2}(m_{D}(T))^2}{2}-\frac{A_{0}(m_{D}(T))}{12}, \beta_{2}=\frac{A_{2}(m_{D}(T))^3}{6}, \beta_{3}=\frac{A_{0}}{2}-A_{2}(m_{D}(T))  \label{pott}.
\end{multline}
The values of constants $A_{0}$, $A_{1}$, $A_{2}$ and $m_{D}(T)$  are not same for every quark-antiquark structure. In the potential model, the coulombic part is because of one gluon exchange. Linear plus quadratic parts of potential model are because of confinement.

\subsection*{2.2. Temperature dependent potential model for hybrid mesons}

Potential model for CM mesons (defined in eq.\ref{pot}) is improved for hybrids as:
\begin{equation}
U(r,T)= \frac{-\beta_{0}}{r}+\beta_{1}r-\beta_{2} r^{2} + \beta_{3} + A (1 - C r^2). \label{hp}
\end{equation}

In the present potential model, an additional term of the form $(A (1 - C r^2))$ is introduced to account for the potential energy difference between conventional(CM) and hybrid (HM) meson states. The constants $A$ and $C$ are determined through fitting the $(A (1 - C r^2))$ with the energy differences ($\upsilon_{d}$) among ground and excited potentials based on lattice simulation results reported in Fig. 3 of Ref.\cite{Morningstar}. The optimal agreement with the lattice data is achieved for $A=1.40498$ GeV and $C=0.016824 GeV^3$ \cite{2024Akber} .

\section* {3. Solution of Radial Schrodinger Equation (SE) and Energy of Mesons by Power Series Method}
\subsection* {3.1 Solution of SE for Conventional Mesons}
Solution of radial SE is stated in Refs.\cite{Kumar,abushadey19} in the following form as:
\begin{equation}
Q(r)=e^{({-\alpha r^{2} - \beta r})}F(r), \label{E2}
\end{equation}
 $F(r)$ is assumed in the form of power series such that:
\begin{equation}
F(r)=\Sigma_{n=0}^{\infty}a_{n} r^{\frac{3n}{2}+L}. \label{E3}
\end{equation}
This form of $F(r)$ is used in Ref.\cite{abushadey19} to calculate the non-degenerate energy eigen states. Here, $a_{n}$ is the expansion coefficient. $\alpha$, $\beta$ are stated below in the present section and their values rely on the potential model.

 By putting the potential model from eq.(4) and power series solutions from eqs.(7,8) in the eq.(\ref{E1}), the following equations are obtained as in Ref. \cite{{Innying2021}}.

 \begin{multline}
\Sigma_{n=0}^{\infty}a_{n}\bigg[\bigg((2n+K)(2n+K-1)+2(2n+K)-K(K+1)\bigg)r^{2n+K-2}+\bigg(-2\beta(2n+K)+(P-2\beta)\bigg) r^{2n+K-1}\\+\bigg(-4\alpha(2n+K)+\epsilon+\beta^{2}-6\alpha\bigg)r^{2n+K}+\bigg(4\alpha\beta-Q\bigg)r^{2n+K+1}+\bigg(4\alpha^{2}+S\bigg)r^{2n+K+2} \bigg]=0
\end{multline}
After equalizing coefficients of exponents of r to zero, the following relations are obtained \cite{{Innying2021}}:
\begin{equation}
\beta = \frac{P}{4n+2K+2}, \label{alpha}
\end{equation}

\begin{equation}
\alpha= \frac{ \sqrt{-S}}{2}, \label{beta}
\end{equation}

\begin{equation}
\epsilon = 2\alpha(4n+2K+3)-\beta^2, \label{c}
\end{equation}

 \begin{multline}
E=\sqrt{\frac{-h^2(A_{2}m^3_{D}(T))}{12\mu}}(3n+2K+3)-\frac{2\mu}{h^2}\bigg(A_{1}-A_{2}+\frac{A_{0}}{m_{D}(T)}\bigg)^{2}\bigg(3n+2K+2\bigg)^{-2}+\\ \frac{A_{0}}{2}-A_{2}m_{D}(T)     \label{E}
\end{multline}

\subsection* {3.2 Solution of SE for Hybrid Mesons}
 In order to compute the properties of hybrid mesons, solution of the SE for the potential model defined in eq.(\ref{hp}) can be assumed as:
\begin{equation}
Q(r)=e^{({-\alpha' r^{2} - \beta' r})}F(r). \label{E2}
\end{equation}
Where $\alpha'$, $\beta'$ are constants and calculated below in present section.

With the modified potential for hybrid mesons defined in eq.(\ref{hp}) and power series solutions from eqs.(8,14), SE can be defined as:
\begin{multline}
\Sigma_{n=0}^{\infty}a_{n}\bigg[\bigg(\frac{(2n+K)}{2}\frac{(2n+K-1)}{2}+2\frac{(3n+K)}{2}-K(K+1)\bigg)r^{2n+K-2}+\bigg(-\beta'(3n+2K)+(P-2\beta')\bigg) r^{2n+K-1}\\+\bigg(-2\alpha'(3n+2K)+\epsilon+\beta'^{2}-6\alpha'\bigg)r^{2n+K}+\bigg(4\alpha'\beta'-Q\bigg)r^{2n+K+1}+\bigg(4\alpha^{2}+S\bigg)r^{2n+K+2} P. \bigg]=0
\end{multline}

After putting the coefficients of exponents of r to zero, following relations are obtained:
\begin{equation}
\epsilon = 2\alpha'(3n+2K+3)-\beta'^2,
\end{equation}
\begin{equation}
\alpha'=\sqrt\frac{-s)}{2}, \label{alphah}
\end{equation}
\begin{equation}
\beta'=\frac{P}{3N+2K+2}, \label{betah}
\end{equation}
\begin{multline}
E =\sqrt{\frac{-h^2(A_{2}m^3_{D}(T)+AC)}{12\mu}}(3n+2K+3)-\frac{2\mu}{h^2}\bigg(A_{1}-A_{2}+\frac{A_{0}}{m_{D}(T)}\bigg)^{2}\bigg(3n+2K+2\bigg)^{-2}+\\ \frac{A_{0}}{2}-A_{2}m_{D}(T)+A. \label{EH}
\end{multline}

\section*{4. Parameters of Potential Models and Mass spectra of mesons}
\subsection*{4.1. Conventional mesons}
Mass of meson can be calculated by adding the masses of quark-antiquark in the energy, E (stated in eq.(\ref{E})), i.e;
\begin{multline}
M = m_{q} +m_{\overline{q}} + +\sqrt{\frac{-h^2(A_{2}m^3_{D}(T))}{12\mu}}(3n+2K+3)-\frac{2\mu}{h^2}\bigg(A_{1}-A_{2}+\frac{A_{0}}{m_{D}(T)}\bigg)^{2}\bigg(3n+2K+2\bigg)^{-2}+\\ \frac{A_{0}}{2}-A_{2}m_{D}(T). . \label{fitt}
\end{multline}

 By using above equation, values of constants ($A_{0}, A_{1}, A_{2}$) can be found by putting $M$ equal to the experimental mass of a specific state of meson ($c\overline{c}$, $b\overline{b}$).
For charmonium, three expressions are found by putting the experimental mass values of $1S$ and $1P$ and $1D$ states in eq.(\ref{fitt}). All three expressions are simultaneously solved and got the values of $A_{0}$ and $A_{1}$ and $A_{2}$.
For bottomonium, same methodology is used to get the value of $A_{0}$, $A_{1}$, and $A_{2}$  by considering the experimental data for mass of $1S$ and $3P$  and $3D$ states.

\subsection*{4.2. Hybrid mesons}

Mass of hybrid mesons is found by the following equation:
\begin{multline}
M =  m_{q} +m_{\overline{q}}+\sqrt{\frac{-h^2(A_{2}m^3_{D}(T)+AC)}{12\mu}}(3n+2K+3)-\frac{2\mu}{h^2}\bigg(A_{1}-A_{2}+\frac{A_{0}}{m_{D}(T)}\bigg)^{2}\bigg(3n+2K+2\bigg)^{-2}+\\ \frac{A_{0}}{2}-A_{2}m_{D}(T)+A \label{mass}
\end{multline}

Here, $\alpha'$ and $\beta'$ are found from eqs.(\ref{alphah},\ref{betah}) where $A_{0},A_{1}$, and $A_{2}$ are the constants of the conventional meson potential model. By using the calculated values of all constants, masses are obtained for ground and excited states of hybrid mesons by taking distinct values of $n,L$.

\section*{5. Results and Discussion}

In the recent study, temperature dependence on the masses of conventional charmonium ($c\overline{c}$), bottomonium ($b\overline{b}$), hybrid charmonium and hybrid bottomonium is studied by employing the power series method. The masses are calculated for different values of the temperature dependent debye mass parameter $m_D(T)$. The calculated mass values for charmonium and bottomonium are reported in Tables (1,2) along with the experimental and other's theoretical calculated masses. From Tables (1,2), It is observed that calculated masses of charmonium and bottomonium show the best agreement with experimental data. The agreement of results with experimental data enhance the validity of power series method and non-relativistic approximation. Masses of hybrid mesons are reported in Tables (3,4) for $m_{D}(T) = 1.52, 1.58, 1.62$ GeV. Masses of hybrid mesons, calculated by others with different methods, are reported in last three columns of Tables (3,4).

Variation in masses for different values of Debye mass $m_{D}(T)$ for $1S$, $1P$, and $1D$ states of conventional and hybrid mesons are plotted in Figs. (1,2). From these figures, it is observed that the calculated masses of hybrid mesons are higher than those of conventional mesons. The calculated values of masses show authentic agreement with other's calculated non-relativistic and relativistic theoretical results. The dependence of masses on the debye mass parameter ($m_{D}(T)$) lead to shifts in energy levels. It is observed that the increase in the Debye mass parameter generally move to higher binding energies and slightly larger meson masses.

\begin{figure}[htbp]
\centering
\begin{tabular}{ccc}
\centering
\includegraphics[width=0.4\textwidth]{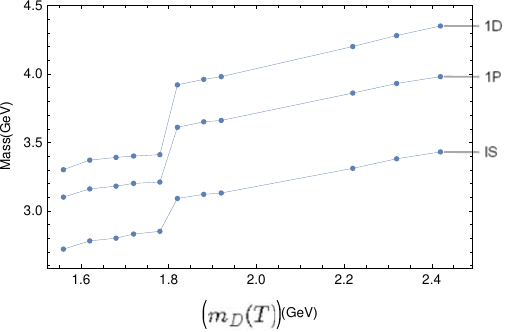}
\hspace{0.1\textwidth}
\includegraphics[width=0.4\textwidth]{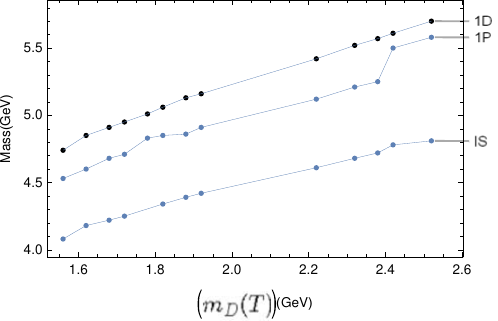}
\hspace{0.1\textwidth}
\end{tabular}
\caption{Variation 0f caharmonium mass with $m_D(T)$. Left side figure is for conventional charmonium and right side figure is for hybrid charmonium.}
\end{figure}

\begin{figure}[htbp]
\centering
\begin{tabular}{ccc}
\centering
\includegraphics[width=0.4\textwidth]{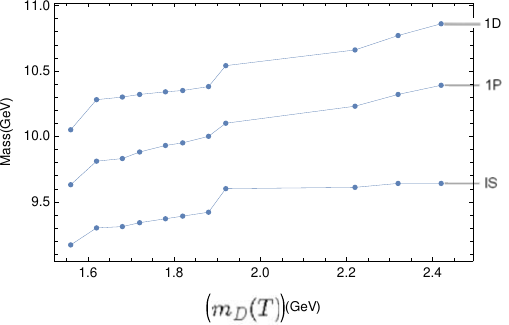}
\hspace{0.1\textwidth}
\includegraphics[width=0.4\textwidth]{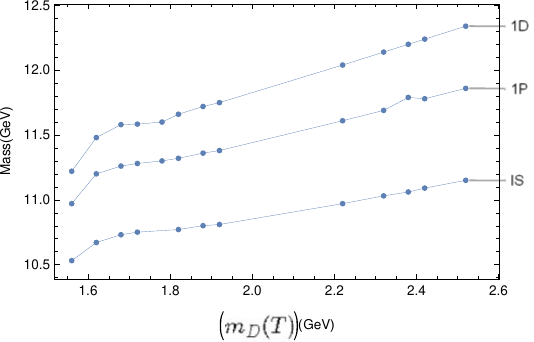}
\hspace{0.1\textwidth}
\end{tabular}
\caption{Variation 0f botomonium mass with $m_D(T)$. Left side figure is for conventional botomonium and right side figure is for hybrid botomonium.}
\end{figure}

Relative errors are calculated for the states whose experimentally calculated mass is available. With $m_D(T)= 1.56 $ GeV, maximum relative error for mass of S, P, D states of charmonium is calculated as 0.057, 0.092, 0.125 respectively; with $m_D(T)= 1.62 $ GeV, maximum relative error for mass of S, P, D states is 0.03, 0.07, 0.106 respectively; with $m_D(T)= 1.68 $ GeV, maximum relative error for mass of S, P, D states is 0.031, 0.06, 0.10 respectively. \\

Similary, maximum relative error for mass of S, P, D states of bottomonium is calculated as 0.03, 0.0284, 0.0062 respectively for $m_D(T)= 1.56 $ GeV; with $m_D(T)= 1.62 $ GeV, maximum relative error for mass of S, P, D states is 0.016, 0.019, 0.0033 respectively; with $m_D(T)= 1.68 $ GeV, maximum relative error for mass of S, P, D states is 0.015, 0.0148, 0.00036 respectively. From these results, it is concluded that $m_{D}(T)$= 1.68 GeV provides the best agreement with experimental data for charmonium and bottomonium mesons.

\begin{table}[h!]
\scriptsize
\begin{center}
\caption{\label{Table1}Mass spectra of $c\overline{c}$ in GeV with $A_{0}$=1.90$\textrm{GeV}^{3}$ and $A_{1}$ = 2.90 $\textrm{GeV}^{2}$ and $A_{2}$= -0.07$\textrm{GeV}$}

\begin{tabular}{ |c|c|c|c|c|c|c|c|c| }
 \hline
 \multicolumn{4}{|c|}{Our calculated mass}&\multicolumn{4}{|c|}{Other calculated mass}&{Mass calculated by experimentalists}\\  \hline
 States & $m_{D}(T)$= 1.56& $m_{D}(T)$=1.62&$m_{D}(T)$=1.68 &\cite{2023Inyang}&\cite{abushadey19}&\cite{Nosheen11}&\cite{Barnes05}&Exp \cite{pdg23}\\ \hline
 1S     & 2.92&2.98&3.00 &3.096& 3.068    & 3.096   &     3.09&   3.0969 $\pm$ 0.000006 \\
 2S     & 3.61&3.65&3.69 &3.686& 3.686    & 3.686&    3.672&    3.6861 $\pm$ 0.000025   \\
 3S     &3.95&3.97& 3.99&4.040& 4.040    &   4.0716&   4.072& -\\
 4S     & 4.17&4.23&4.28&4.263 & 4.269   &  4.406  &   4.406& - \\
 1P     & 3.10&3.16&3.18&3.525  &  3.255   &   3.4245&   3.424  &     3.41475 $\pm$ 0.00031 \\
 2P     & 4.03&4.07&4.11&3.772 & 3.779    &   3.8523 &  3.852& -\\
 3P     & 4.10&4.31&4.21 &-&    -    &    4.2017&   4.202&  - \\
 4P     & 4.32&4.39&4.45&-&     -     &   4.5092&   -    &   -\\
 1D     & 3.30&3.37&3.39&3.770 &  3.504      & 3.7850&  3.785 &        3.77313 $\pm$ 0.00035 \\
 2D     & 4.02&4.07&4.12&4.159 &    -     &    4.1415 & 4.142 &         4.159 $\pm$ 0.00021 \\
 3D     & 4.31&4.31&4.37&-&     -     &     4.4547 & -    & -\\

 \hline
\end{tabular}
\end{center}
\end{table}
\begin{table}[h!]
\begin{center}
\scriptsize
\caption{\label{Table2} Mass spectra of $b\overline{b}$ in GeV with $A_{0}$=1.90$\textrm{GeV}^{3}$ and $A_{1}$ = 2.90 $\textrm{GeV}^{2}$ and $A_{2}$= -0.07$\textrm{GeV}$}
\begin{tabular}{ |c|c|c|c|c|c|c|c|c|c|}
 \hline
 \multicolumn{4}{|c|}{Our calculated mass}&\multicolumn{4}{|c|}{Other calculated mass}&{Mass calculated by experimentalists}\\  \hline
 States & $m_{D}(T)$=1.56&$m_{D}(T)$=1.62&$m_{D}(T)$=1.68&\cite{2023Inyang}  &  \cite{abushadey19}  & \cite{Nosheen17} & \cite{14110585}&  Exp \cite{pdg23}  \\ \hline
 1S     & 9.17&9.30&9.31 &9.460& 9.46  &     9.5299   & 9.459&  9.4603 $\pm$ 0.00026  \\
 2S     & 9.78&9.98&10.00&10.023  & 10.023&   10.010  &   10.004& 10.023 $\pm$ 0.00031 \\
 3S     & 10.09&10.38&10.39&10.355 & 10.585&   10.295  &   10.354& 10.3552 $\pm$ 0.0005   \\
 4S     & 10.34&10.68&10.71&10.580& 11.148&   10.5244 &    10.663 & 10.5794 $\pm$ 0.0012  \\

 1P     & 9.63&9.81&9.83&9.898& 9.8354     &9.9326 &    9.896 &   9.91221 $\pm$ 0.00026  \\
 2P     & 10.00&10.25&10.27&10.256 & 10.398 &   10.2245&    10.261&    10.26865 $\pm$ 0.00022 \\
 3P     & 10.33&10.65&10.68&-&   -     &     10.4585&   10.549&    10.524 $\pm$ 0.0008 \\
 4P     & 10.49&10.87&10.91& -&-      &     10.6627&      10.797&   -   \\
 1D     & 10.10&10.13&10.16&10.164& 10.210      &     10.1389&10.155&    10.1637 $\pm$0.0014  \\
 2D     & 10.18&10.47&10.56&10.306& -         &    10.3799& 10.455&      - \\
 3D     & 10.46&10.83&10.56& -&-          &   10.5892& 10.711&      - \\
 4D     & 10.62&11.06&11.10& -&-           &   10.7782& 10.939&     - \\ \hline
\end{tabular}
\end{center}
\end{table}
\begin{table}[h!]
\scriptsize
\begin{center}
\caption{\label{Table 3} Mass spectra of hybrid Charmonium  in (GeV).}
\begin{tabular}{ |c|c|c|c|c|c||c|c|c|c|c|c|}
\hline
\multicolumn{5}{|c|}{Our calculated mass}&\multicolumn{3}{|c|}{Other calculated mass}\\  \hline
 States & $m_{D}(T)$=1.56&$m_{D}(T)$=1.62 &$m_{D}(T)$=1.68&$m_{D}(T)$    =1.78&\cite{2024Akber}& NR\cite{Nosheen2014} & Rel. \cite{Nosheen2014}    \\ \hline
1S     & 4.08&4.18&4.22&4.30&4.29 &  4.1063      &    4.1707  \\
2S     & 4.65&4.47&4.82&4.92&4.55 &  4.4084     &     4.4837  \\
3S     & 4.86&5.01&5.09 &5.20&4.80&  4.6855    &      4.7614  \\
4S     &  4.98   & 5.15&5.23& 5.29&   5.05&    4.94&5.01\\

1P     & 4.53&5.02&4.68&4.83 &4.46&  4.2464    &      4.3203  \\
2P     & 4.80&5.12& 5.14&5.18&4.71&   4.5264    &      4.6070 \\
3P     & 4.95&5.10&5.19&5.32&4.97&  4.7875     &     4.8659   \\
4P     & 5.15 &5.23&5.34&5.49&5.22&  5.0338     &     5.1034  \\

1D     & 4.74&4.85 & 4.91&5.01&4.63&  4.4232   &       4.4320 \\
2D     & 4.90&5.05 &5.13&5.25&4.88&  4.6955    &      4.6892\\
3D     & 5.10 &5.19&  5.29&5.44&5.14&  4.9402 &         4.966\\
4D     & 5.11&5.31 &5.60 &5.60&5.39& 5.1912    &      5.2034  \\
 \hline
\end{tabular}
\end{center}
\end{table}

\begin{table}[h!]
\scriptsize
\begin{center}
\caption{\label{Table 4} Mass spectra hybrid botmonium in (GeV). }
\begin{tabular}{ |c|c|c|c|c|c|c|c|}
\hline
\multicolumn{5}{|c|}{Our calculated mass}&\multicolumn{3}{|c|}{Other calculated mass}\\  \hline
 States &$m_{D}(T)$=1.56&$m_{D}(T)$=1.62&$m_{D}(T)$=1.68&$m_{D}(T)$=1.78& \cite{2024Akber}   & NR \cite{Nosheen17}& Rel. \cite{Nosheen17}   \\ \hline
1S     & 10.53&10.76&10.73&10.57 &10.80&       10.7747&     10.8079\\
2S     & 11.11&11.34&11.43&11.37 &11.17&         10.9211&     10.928\\
3S     & 11.40&11.71&11.83&11.80 &11.53&       11.0664&     11.048\\
4S     & 11.90&12.01&12.16&12.14 & 11.90&      11.2086&     11.1662\\

1P     & 10.97&11.18&11.26&11.18 &11.05&   10.8366&      10.8569 \\
2P     & 11.32&11.60&11.71&11.67&11.41&      10.9857&      10.9856 \\
3P     & 11.55&11.99&12.13&12.12&11.77&   11.1372&      11.1097 \\
4P     & 12.10&12.19&12.37&12.35 &12.14&     11.2795&      11.2293 \\

1D     & 11.22&11.48&11.58&11.53 &11.29&    10.9063&      10.9127 \\
2D     & 11.48&11.81&11.71&11.67 &11.65  &     11.0591&      11.0415 \\
3D     & 11.69&12.16&12.32 & 12.02&    12.16&   11.2054&      11.165 \\
4D     & 12.12&12.36&12.56&12.54  &12.38&     11.3463&      11.2837 \\
 \hline
\end{tabular}
\end{center}
\end{table}
\section*{6. Conclusion}
In this paper, mass spectra of conventional charmonium, bottomonium, hybrid charmonium and hybrid bottomonium states have been studied through the power series method by using the temperature-dependent potential model. It is noted that the increase in Debye mass increases the energy eigenvalues. Debye mass is directly proportional to temperature\cite{NP21}. Therefore, it is concluded that mass of conventional (CM) and hybrid (HM) mesons increase with the increase of temperature. The agreement of the calculated masses with the experimental data proves the success of power series method and validity of non-relativistic Schrodinger equation for heavy quark-antiquark systems.

The predicted masses of hybrid charmonium and hybrid bottomonium show a clear difference of hybrids from conventional states, reflecting the contribution of excited gluonic fields. Furthermore, the sensitivity of masses of hybrid mesons with the variations in the debye mass parameter highlights the importance of precise parameter selection in meson spectroscopy studies.

\end{document}